\documentclass[aps,pre,twocolumn,superscriptaddress,showpacs]{revtex4}
\usepackage{amsmath,amssymb,graphicx,color}

\begin{document}
%\draft
%\twocolumn[\hsize\textwidth\columnwidth\hsize\csname@twocolumnfalse\endcsname
\title{Comparison of free energy estimators and their dependence on dissipated work}

\author{Seongjin Kim}
\author{Yong Woon Kim}
\affiliation{Graduate School of Nanoscience and Technology, Korea Advanced Institute of Science and Technology, Daejeon 305-701, Korea}
\author{Peter Talkner}
\affiliation{Institute of Physics, University of Augsburg, Universit\"{a}tsstrasse 1, D-86135 Augsburg, Germany}
\author{Juyeon Yi}
\affiliation{Department of Physics, Pusan National University, Busan 609-735, Korea}

\date{\today}

\begin{abstract}
The estimate of free energy changes based on Bennett's acceptance ratio method is examined
in several limiting cases and compared with other estimates based on the Jarzynski equality and on the Crooks relation.
While the absolute amount of dissipated work, defined as the surplus of average work over the free energy difference,
limits the practical applicability of Jarzynski's and Crooks' methods, the reliability of Bennett's approach is restricted by
the difference of the dissipated works in the forward and the backward process.
We illustrate these points by considering a Gaussian chain and a hairpin chain
which both are extended during the forward and accordingly compressed during the backward protocol.
The reliability of the Crooks relation predominantly depends on the sample size;
for the Jarzynski estimator the slowness of the work protocol is crucial,
and the Bennett method is shown to give precise estimates irrespective of the pulling speed and sample size
as long as the dissipated works are the same for the forward and the backward process as it is the case for Gaussian work distributions.
With an increasing dissipated work difference the Bennett estimator also acquires a bias which increases roughly in proportion to this difference.
A substantial simplification of the Bennett estimator is provided by the 1/2-formula which expresses
the free energy difference by the algebraic average of the Jarzynski estimates for the forward and the backward processes.
It agrees with the Bennett estimate in all cases when the Jarzynski and the Crooks estimates fail to give reliable results.
\end{abstract}


\maketitle

\section{Introduction}
In studying the thermodynamic state of a physical system, the free energy $F$ is a quantity of fundamental importance.
It describes the equilibrium properties of systems that may exchange energy with their environments.
Formally, it is related to the internal energy, $U$, by a Legendre transform
$F=U-TS$, where $T$ is the temperature and $S$ the entropy.
The free energy is a state function and hence,
for any process connecting two equilibrium states, the respective change of the free energy
$\Delta F=\Delta U -T\Delta S$, solely depends on the final and initial states without regard to the particular process connecting them.
In contrast, the work $w$ done on the system and the heat $Q$ exchanged with the environment are process-dependent.
Yet, their sum yielding the change in internal energy, $\Delta U =w+Q$,  does not depend  on the details of the path connecting the final with the initial state.

Recently, Jarzynski found a relation between the path dependent work and the path independent free energy change
in terms of the following sum rule~\cite{jarzynski},
\begin{equation}\label{jarzynski}
\int_{-\infty}^{\infty}dw p(w) e^{-\beta w}=\langle e^{-\beta w}\rangle = e^{-\beta \Delta F},
\end{equation}
where $p(w)$ denotes the probability density function (PDF) of the work that
is performed on the system. The process from which this work results, starts out in a state of thermal equilibrium at temperature $T=(k_B \beta)^{-1}$, and is induced by the action of forces, or more generally by changes of  parameters characterizing the Hamiltonian of the considered system. These parameter changes are supposed to follow a specific protocol, on the details of which the work PDF will depend in general.

In Eq.~(\ref{jarzynski}), $\Delta F =F_{f}-F_{i}$,  denotes the difference between the free energies $F_i$ and $F_f$ of the initial thermal equilibrium state and the thermal equilibrium state of the system with the final parameter values, respectively. Both equilibrium states are at the same temperature $T$. In general, the second equilibrium state differs from the actual state that is reached at the end of the protocol. In principle, this ``associated thermal equilibrium'' will be approached when the system stays in contact  with a heat bath at temperature $T$ upon completion of the protocol.
As a difference of a state function, $\Delta F$ depends on the initial and final parameter values but is independent of the details of the protocol.

The random nature of the work manifested in the PDF $p(w)$ is a consequence of the inherent randomness of the thermal initial state and additionally also due to a possible randomness of the dynamics, be it of quantum or classical, stochastic nature. In the latter case, randomness and dissipation must be properly balanced by fluctuation dissipation relations eventually imposing thermal equilibrium at the initial inverse temperature $\beta$ at constant parameter values.
Any application of the Jarzynski equality (\ref{jarzynski}) to experiments hence requires repeating experiments with the same protocol many times in order to generate a sufficient statistics. The same requirement to generate sufficiently many data must also be met in numerical implementations of the Jarzynski equality aiming at the determination of free energy changes.

The feasibility of this scheme with the goal to determine the free energy change $\Delta F$ was demonstrated in various experimental systems such as for single molecules~\cite{sm1,sm2} and classical oscillators~\cite{pen1,pen2}.
Yet the practical applicability of the Jarzynski equality is severely limited because the estimate of the exponential work average strongly relies on how well the tail of the work distribution with $w < \Delta F$ is sampled~\cite{jarzynski2}. In general the exponential average $\langle e^{-\beta w} \rangle$ gives rise to a bias of the Jarzynski free energy estimate~\cite{jarzynski2} .
It is now well recognized that the best convergence can be obtained  from slow protocols associated with
small dissipation such that $\langle w\rangle \approx \Delta F$, i.e., for almost reversible processes
in which the system passes through a succession of quasi-equilibrium states~\cite{fox}.
For fast switching, the finite sampling error becomes substantial both concerning the bias and the variance. The respective behavior in the large sampling limit was investigated in several studies~\cite{bias1,bias2,zucker1,zucker2}.

An alternative strategy to determine free energy differences is based on the Crooks relation~\cite{crooks},
\begin{equation}\label{crooks}
p_{f}(w)=e^{-\beta (\Delta F -w)}p_{b}(-w)
\end{equation}
which allows to infer $\Delta F$ without the average process required in Eq.(\ref{jarzynski}).
Here the two work PDFs $p_f(w)$ and $p_b(w)$ refer to the original, or forward ($f$) protocol and to the backward protocol ($b$)
which is started in the associated equilibrium state, i.e., at the initial temperature of the forward process and at those parameter values
that have finally been reached in the forward process and retraces its parameter values.
%are obtained from bi-directional work: Say, $p_{f}(w)$ from forward protocol,
%and $p_{b}(w)$ from its reversed protocol.
The free energy change can be read off from a plot displaying the functions $p_f(w)$ and $p_b(-w)$
as the work value at which the two distributions cross each other.
%Note that $p_{f}(\Delta F)=p_{b}(-\Delta F)$, indicating
%that the cross point between $p_{f}(w)$ and $p_{b}(-w)$ corresponds to $\Delta F$.
A reliable estimate of the crossing point requires sufficient sampling of work with $w < \Delta F$
for the forward and respectively with $w < -\Delta F$ for the backward process.

It is worthwhile mentioning here that the {\it average} work is never smaller than the free energy change, i.e. $\langle w\rangle- \Delta F \geq 0$,
as a consequence of Jensen's inequality, stating that $\langle e^{-\beta w}\rangle \geq e^{-\beta \langle w\rangle}$. Hence, the work applied to the system in an isothermal process is at least as large as the change of free energy, in accordance with the second law of thermodynamics. The amount of work that is equal to the free energy change can be applied to the system in an isothermal reversible process, while any surplus $\langle w \rangle -\Delta F$ is ``dissipated work''. The dissipated work can also be interpreted in terms of the entropy change of the total system including the heat-bath, $\langle w\rangle - \Delta F = T\Delta S_{tot}$~\cite{entropy2}, which, again is positive in accordance with the second law.
%Interpreting $\langle w\rangle - \Delta F = T\Delta S_{tot}$ with $\Delta S_{tot}$ being the total entropy production in system plus reservoir, its positivity is an indication of the second law of thermodynamics~\cite{entropy2}.} This average also characterizes the so-called ``dissipated work''.
For rapid protocols, the dissipated work becomes large and therefore those realizations of the process with $w<\Delta F$ may become extremely unlikely, such that even large samples obtained from experiments or numerical investigations may result in
%leading to
an unpopulated no man's land between the forward and the backward work PDFs and,
hence, in an
unreliable estimate of the free energy change based on Crooks' crossing criterion.

Dividing both sides of the Crooks relation~(\ref{crooks}) by the factor of
%We note here a modified version of the Crooks relation given by
$1+e^{\beta(w-\Delta F)}$, one obtains
\begin{equation}\label{bennett}
\int_{-\infty}^{\infty}dw \frac{p_{f}(w)}{1+e^{\beta (w-\Delta F)}}=\int_{-\infty}^{\infty}dw
\frac{p_{b}(-w)}{1+e^{-\beta(w-\Delta F)}}.
\end{equation}
%Multiplying the both sides of Eq.~(\ref{crooks}) with the fermi function-like factor, $1+e^{\beta(w-\Delta F)}$ and integration over $w$ leads to Eq.~(\ref{bennett}).
This equation was originally suggested by Bennett ~\cite{bennett} as an efficient basis to estimate partition function ratios by means of Monte Carlo sampling. In Bennett's derivation, the Fermi-function-like weights in Eq.~(\ref{bennett}) resulted as acceptance ratios from the requirement of minimal variance of the partition function estimator in the large sample size limit. From the Gaussian assumption which this argument underlies one would be let to believe that
the Bennett estimator yields minimal variance only if the overlap region of the forward and backward PDFs $p_f(w)$ and $p_b(-w)$ is substantially populated. However, Shirts et al.~\cite{shirts} demonstrated
that Bennett's acceptance ratio method always yields
%can be understood as
a maximum likelihood
estimate of the free energy change by the use of forward and backward data.
As such, it allows to extract $\Delta F$ with the smallest variance compared to any other free energy estimator, even if the two PDFs, $p_{f}(w)$ and $p_{b}(-w)$ do not overlap. Therefore, %$\Delta F$
the estimation of the free energy change from Eq.~(3) is less restricted than the Jarzynski method and Crooks' crossing criterion as has been confirmed in various recent studies~\cite{P,Y,N}.
%{\bf Recent studies report that the Bennett estimation outperforms other estimators~\cite{P,Y,N}, and yet}
%its correctness is limited by the difference between dissipated works in the forward and the backward processes as we shall demonstrate.

In this work, we investigate the statistical behavior of the free energy change estimation
by the Bennett method, discuss its limitations imposed by the difference of the amounts of dissipated work in the forward and backward protocol, and compare it with the methods proposed by Jarzynski and Crooks. In our discussion we lay %with
particular
emphasis %laid
on the practical limitations resulting from the nonequilibrium nature that
is imposed on the system by time dependent perturbations.
We begin in Section II with a brief review on how the  dissipated work and the time asymmetry constrain the Jarzynski and Crooks method.
In the following Section III, we shortly review the maximum likelihood argument \cite{shirts} leading to  Eq.~(\ref{bennett}) and examine its solutions  in some limiting cases.
In particular when the Jarzynski and Crooks methods are both hampered by large dissipated work,
we show that the Bennett method simplifies to the ``1/2-formula'' expressing the free energy change as the arithmetic mean of
the Jarzynski estimation obtained from the forward and the backward process.
In Sec. IV, we take as an example a Gaussian chain, and consider processes in which work is done by
extending or compressing the chain at a constant speed.
Reliable estimations of $\Delta F$ based on Crooks' crossing criterion or the Jarzynski estimator do not exist if the dissipated work becomes
large. We present regions in a pulling-speed versus sample-size plane, in which the errors of these estimators are smaller than $k_{B}T$.
For the Gaussian chain, the dissipated works in the forward and backward processes are the same, %accompanied by equal amounts of dissipative work,
and, as a consequence the Bennett method gives precise estimations of $\Delta F$.
In Sec. VI, we consider a chain of monomers interacting via pairing potentials to form a hairpin-like structure.
Keeping one end fixed and pulling at the other end with constant speed gives rise to a non-Gaussian work distribution
with different dissipated works for the forward and the backward protocols. The difference increases with growing pulling speed
and leads to a finite bias of the Bennett estimator. The results are summarized in Sec. V.

\section{generalities}

We consider a time dependent perturbation
of a system in terms of a
%applied to a system, and let the
control parameter %of the perturbation, say
$\lambda(t)$ that varies in time $t$ according to a prescribed protocol. The protocol can be performed in bidirectional way. For the forward protocol during a time interval $t\in (t_{i},t_{f})$, the system departs from its initial thermal equilibrium state at temperature $T$ and control parameter $\lambda (t_{i})$ %in thermal equilibrium state with a reservoir of
%temperature $T$,
and reaches a terminal state with $\lambda(t_{f})$, which, in general, is a nonequilibrium state. The work done during this process will be referred to as the forward work. For the backward protocol, the system %is initially
starts out in thermal equilibrium at the same inverse temperature $\beta$ with the control parameter
%perturbation
at $\lambda(t_{f})$. In this case, the parameter retraces its values %{\cancel {in reversed order} {\cblu
in time-reversed order
and finally reaches %varies along the time-reversed path of $\lambda(t)$ and $\lambda(t)$ at the end of the backward protocol reaches
$\lambda(t_{i})$ as terminal parameter value. After repeating infinitely many times %of these two directional
both protocols, two probability distribution functions, $p_{f}(w)$ for the forward and $p_{b}(w)$ for the backward protocol, can be assembled.

%%%%%%%%%%%%%%%%%%%%%%%%%%%%%%%%%%%%%%%%%%%%%%%%%%%%%%%%%%%%%%%%%%%%%%%%%%%%%%%%%%%%%%%%%%%%%%%
\begin{figure}[t]
%\resizebox{7cm}{!}{
\includegraphics[width=.9\columnwidth]{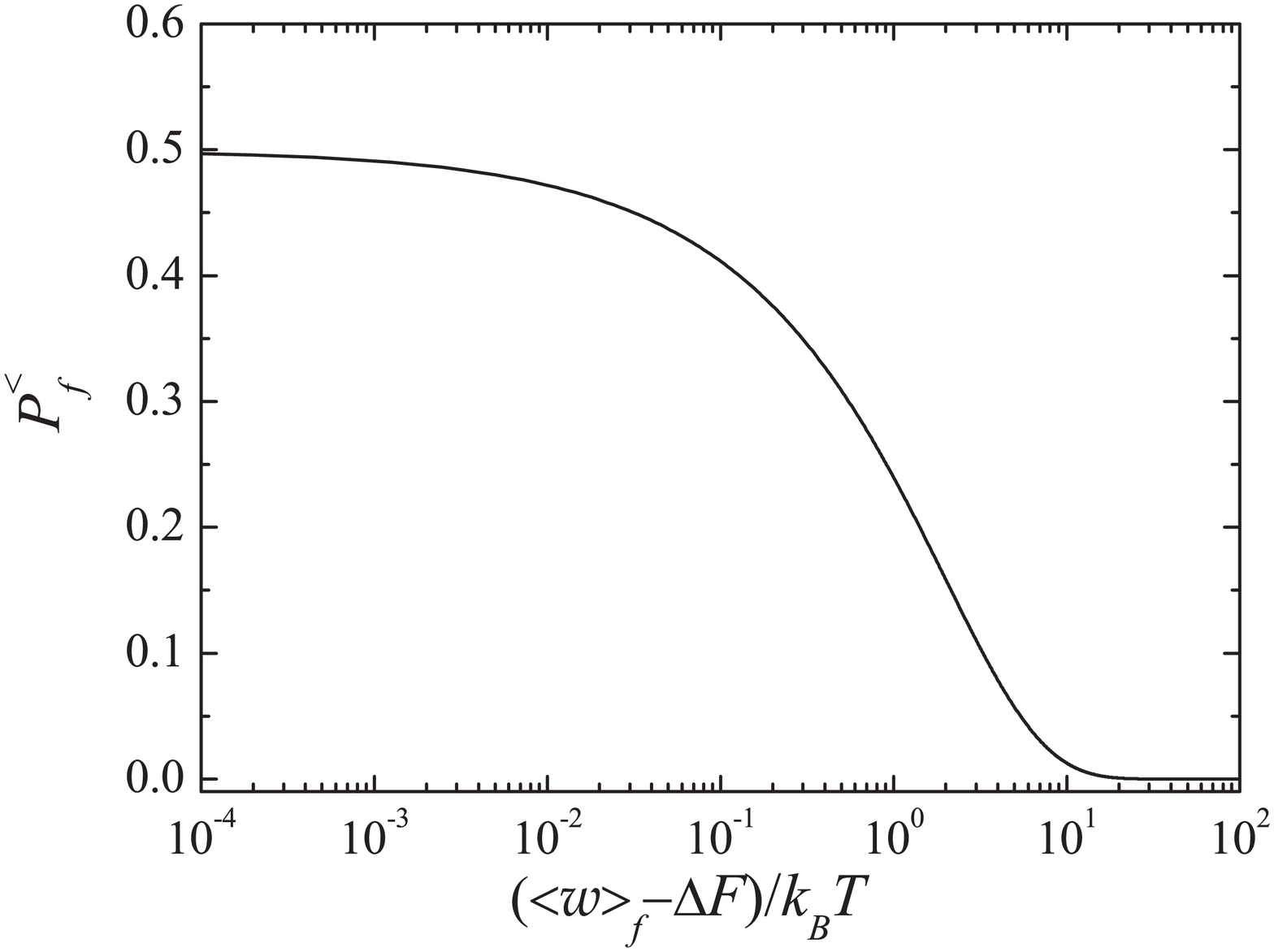}
%}
 \caption{The probability $P^{<}_f$ to obtain a work smaller than the free energy change in a realization of the forward process is displayed  as a function of the dissipated work $\langle w \rangle_f - \Delta F = \beta \sigma^2/2$ for a Gaussian process as given by Eq. (7) . Almost reversible protocols with small dissipated work give a probability $P^{<}_f$ close to the maximum $1/2$. In the opposite limit of large dissipated work $P^{<}_f$ becomes exponentially small.
%approaches is maximum value $1/2$
% Eq. (7) as a function of $\langle w\rangle_f-\Delta F=\beta\sigma^2/2$.
% For weakly dissipative processes ($\beta\sigma \ll 1$), $P^{<}_f$ approaches $1/2$, while for the opposite case, $P^{<}_f$ becomes
% vanishingly small.
}
\label{fig1}
\end{figure}
%%%%%%%%%%%%%%%%%%%%%%%%%%%%%%%%%%%%%%%%%%%%%%%%%%%%%%%%%%%%%%%%%%%%%%%%

As mentioned in the introduction, the probability for observing work less than the free energy change is the crucial factor
for the applicability of the
%in using the
Crooks' crossing criterion. For the forward process this probability is given by
\begin{equation}
P_{f}^{<}\equiv \int_{-\infty}^{\Delta F}dw p_{f}(w),
\end{equation}
and, accordingly, for the backward process, by
\begin{equation}
P_{b}^{<}\equiv \int_{-\infty}^{-\Delta F}dw p_{b}(w).
\end{equation}
If these probabilities are too low to get a sufficient number of realizations of work
%realize the work
below the free energy change within a reasonable total number of experiments, the Crooks' crossing criterion fails to give $\Delta F$.
Moreover, a poor sampling of work data below $\Delta F$ will lead to a substantial underestimation of the exponential average $\langle e^{-\beta w} \rangle$ resulting in a too large Jarzynski estimate of $\Delta F$.

As an example, we consider a Gaussian work PDF $p_f(w)$ for the forward process. As a consequence of the Crooks relation~(\ref{crooks}) the backward work PDF $p_b(w)$ is also Gaussian with the same variance $\sigma^2$. Hence the two PDFs are:  %and backward ($b$) work PDFs
\begin{equation}\label{gauss}
p_{\alpha}(w)=\frac{1}{\sqrt{2\pi\sigma^{2}}}\exp\left[-\frac{(w-\langle w\rangle_{\alpha})^{2}}{2\sigma^{2}}\right], \quad \alpha = f,b\:.
\end{equation}
In this case the free energy change becomes $\Delta F = \langle w\rangle_f -\beta \sigma^{2}/2 = -\langle w \rangle_b + \beta \sigma^2/2$. The probability of finding work less than the free energy change in the forward protocol  then becomes
\begin{equation}\label{negativework}
P_f^{<}
=\frac{1}{2}\text{erfc}\left(\frac{\beta \sigma}{2\sqrt{2}}\right),
\end{equation}
where $\text{erfc}(x)$ denotes the complementary error function, see Fig.~1.
If the process is weakly dissipative, i.e for $\beta(\langle w \rangle_{f}- \Delta F) =(\beta \sigma)^{2}/2 <1$,
the forward work distribution becomes narrow and the argument of the error function is small.
Because of
%Then, since
$\text{erfc}(x)\approx 1 -2x/\sqrt{\pi}$ for $x \ll 1$, the probability $P_f^{<}$ %\approx 1/2$,
approaches the value $1/2$ in the limit of vanishing dissipated work~\cite{chari}.
%implying that on average
In other words, on average, every other measurement probes a work that is less than the free energy change (Fig.~1).
In the opposite limit of a strongly dissipative process
%On the other hand, for a system subject to a strongly dissipative process~(
being characterized by $\langle w\rangle_{f}-\Delta F \gg \beta^{-1}$, %the work PDF becomes disperse for
the work PDFs become broad because then $\beta\sigma \gg 1$.
Using the asymptotic behavior of $\text{erfc}(x)$ for $x\gg 1$, $\text{erfc}(x)\approx e^{-x^2}/(\sqrt{\pi}x)$, we %find
obtain an exponentially small probability
$P_f^{<}\sim \exp(-\beta^{2}\sigma^{2}/8)/(\beta \sigma)$ to find a work in the forward protocol that is less than the free energy change.

%%%%%%%%%%%%%%%%%%%%%%%%%%%%%%%%%%%%%%%%%%%%%%%%%%%%%%%%%%%%%%%%%%%%%%%%%%%%%%%%%%%%%%%%%%%%%%%
\begin{figure}[t]
%\resizebox{7cm}{!}{
\includegraphics[width=.9\columnwidth]{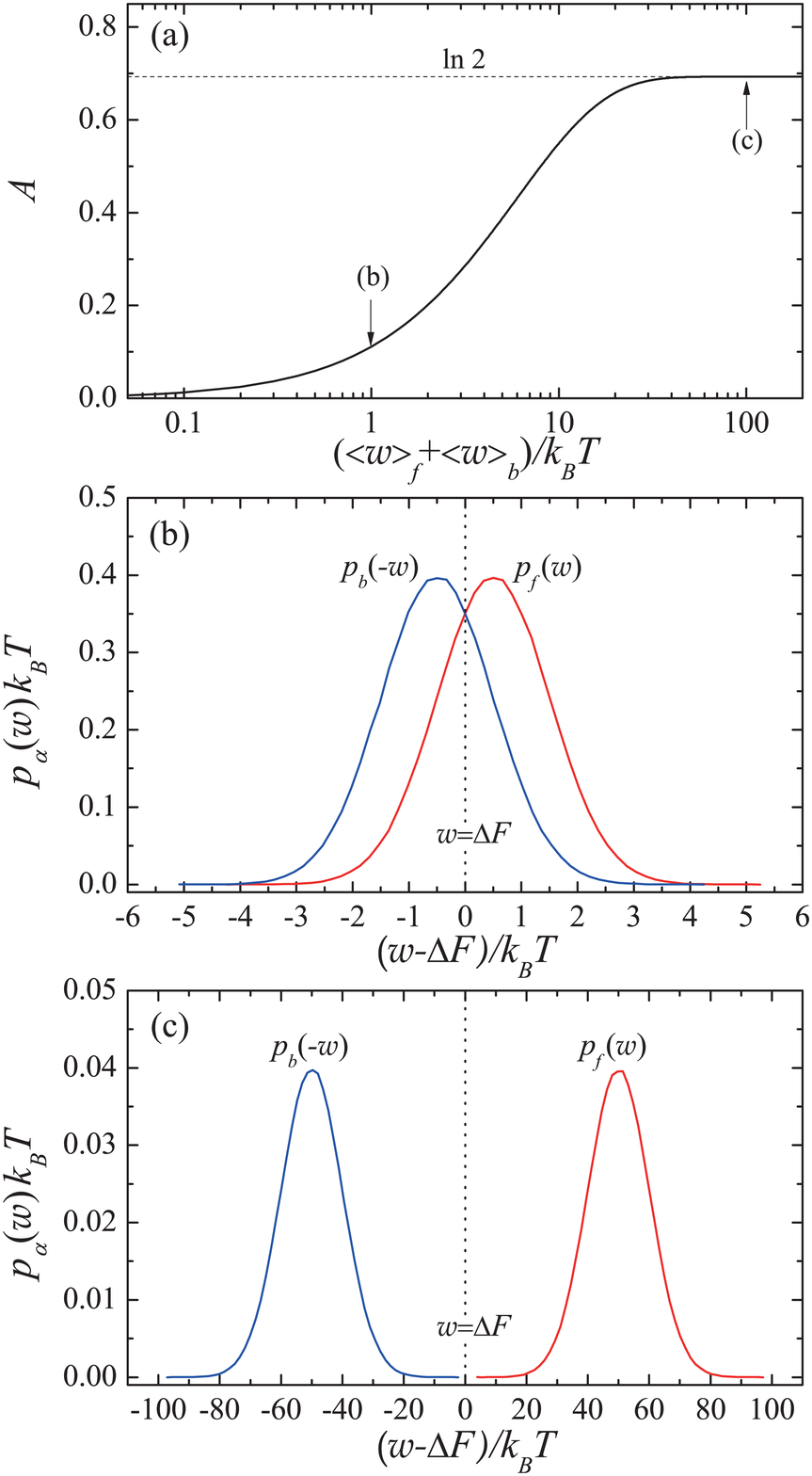}
%}
 \caption{
(Color online)~(a) The time asymmetry $A$ defined by Eq.~(6) for a Gaussian PDF monotonicly varies as a function  of the hysteresis $h=( \langle w \rangle_{f}+ \langle w \rangle_{b})/2=
\beta\sigma^2/2$ from vanishingly small values at small $h$ to a maximal value at large $h$. Note that the hysteresis $h$ gives the average dissipated work for the forward and the backward protocol, and, for a Gaussian work distribution, characterizes the separation of the peaks of the forward and backward work pdfs %as a measure of total dissipation or separation between
$p_f(w)$ and $p_b(-w)$, respectively.
The PDFs at the values indicated by the arrows in panel (a) are displayed in the panels (b) and (c).
(b) For weak dissipation ($(\beta\sigma)^2=1$), the time asymmetry is small ($A\approx 0.1$)
and the work PDFs display a large overlapping area near $\Delta F$.
(c) For strong dissipation ($\beta\sigma^2=100$), the time asymmetry approaches its maximum ($A = \ln 2$)
and the work PDFs are well-separated.
}
\label{fig2}
\end{figure}
%%%%%%%%%%%%%%%%%%%%%%%%%%%%%%%%%%%%%%%%%%%%%%%%%%%%%%%%%%%%%%%%%%%%%%%%

A measure quantifying the difference between the forward and the backward work PDFs $p_f(w)$ and $p_b(-w)$ based on the Jensen-Shannon divergence of the two distributions is given by the so-called
%A related measure to the nonequilibrium entropy production is the
time asymmetry~\cite{feng,J,CH} %of Jensen-Shannon divergence considered
reading
\begin{equation}
A=\frac{1}{2}\left\langle\ln\frac{2}{1+e^{-\beta(w-\Delta F)}}\right\rangle_{f}
+\frac{1}{2}\left\langle\ln\frac{2}{1+e^{-\beta(w+\Delta F)}}\right\rangle_{b},
\label{A}
\end{equation}
where $\langle X(w) \rangle_{f}=\int dw p_{f}(w)X(w)$ and $\langle X(w) \rangle_{b}=
\int dw p_{b}(w)X(w)$ denote averages with respect to %of
the forward and the backward distributions, respectively.  This quantity vanishes when $p_f(w)$ and $p_b(-w)$ agree with each other and otherwise falls  in the range $0\leq A \leq \ln 2$. Therefore, if the measured work values are mostly populated in the close vicinity of the free energy change, $\Delta F$ for the forward protocol and $-\Delta F$ for the backward protocol, the time asymmetry approaches its minimum value $A=0$. In this case, $p_{f}(w)$ and $p_{b}(-w)$ have a large overlap. On the other hand, the time asymmetry reaches its maximum $\ln 2$ when $p_{f}(w)$ and $p_{b}(-w)$ are perfectly separated by a large gap between the  respective regions with  $w\gg \Delta F$ for the forward protocol and
$w \gg -\Delta F$ for the backward protocol.

We exemplify these behaviors for Gaussian work PDFs, given by Eq.~(\ref{gauss}): Figure 2~(a) displays the time asymmetry as a function of %against
the peak separation between $p_{f}(w)$ and $p_{b}(-w)$, that is as a function of the hysteresis $h$ defined as \cite{feng}
\begin{equation}
h = \frac{1}{2}(\langle w\rangle_{f}+\langle w\rangle_{b})\:.
\label{hysteresis}
\end{equation}
For small values of $h$, the time asymmetry is close to zero, and the corresponding PDFs %, for example at $D=1$,
substantially overlap in an
%show a large overlapping
area near $\Delta F$ as displayed in Fig.~2(b) for $\beta h=1/2$. In contrast, when $h$ is large %~($D=100$),
the time asymmetry approaches its maximum, and seemingly, the forward and backward PDFs %the overlap between
$p_{f}(w)$ and $p_{b}(-w)$ no longer overlap with each other %is inappreciable
as exemplified in Fig.~2(c) for $\beta h =50$. Strictly speaking, due to the Crooks relation~(\ref{crooks}) $p_{f}(w)$ and $p_{b}(-w)$ must share the same support and therefore always have a finite overlap, which, however, may become virtually invisible. %{\color{red}\cite{overlap}}. %extremely small.
Little overlap of the forward and the backward work PDFs
indicates that the process involved in the work generation is highly irreversible, and hence, the direction of time flow is unambiguous giving the
information $\ln 2$ of a single bit.  Also, for this case, large dissipated work
is associated with both processes, and the probabilities $P^{<}_{b}$ %~({\bf the area under the curve $p_{b}(-w)$ with $w \geq \Delta F$})
and $P^{<}_{f}$ %~({\bf the area under the curve $p_{f}(w)$ with $w\leq \Delta F$})
to find work values being less than the free energy change become extremely small (Fig.~2(c)).
In summary, the more pronounced is the direction of the time-arrow in a nonequilibrium process, the rarer
it becomes to find events with work smaller than the free energy change
and the more erroneous both estimators by Jarzynski and Crooks turn out to be.

%It is worthwhile to
Finally, we note a relation between the time asymmetry and %nonequilibrium entropy production.
the hysteresis $h$ specifying the average dissipated work of the forward and the backward protocol.
Because the free energy does not change upon completion of a cyclic process,
the total dissipated work agrees with twice the  above introduced work  hysteresis, $2 h$. %, given by
%\begin{equation}\label{hysteresis}
%h=\frac{1}{2} (\langle w\rangle_{f} + \langle w\rangle_{b}).
%\end{equation}
As pointed out in Ref.~\cite{jarzynski2}, the hysteresis gives a rough estimate for the number of measurements $M_{c}$
required for a relatively %{\cancel {correct}} {\cblu
reliable estimation of $\Delta F$ based on the Jarzynski equality,
as $M_{c} \gtrsim e^{\beta h}$. On the other hand the hysteresis is %meanwhile bounded by
related to the time asymmetry by the following inequality~\cite{feng},
\begin{equation}\label{haineq}
e^{\beta h} \geq \frac{1}{2 e^{-A}-1}.
\end{equation}
As a consequence the required number of measurements becomes infinitely large
if the time asymmetry approaches the limiting value, $\ln 2$.
Therefore not only the Jarzynski estimator ceases to work but also Crooks' crossing criterion
fails for processes with unambiguous arrows of time.

\section{Bennett's acceptance ratio method}

As mentioned in the Introduction, Shirts et al. \cite{shirts} obtained the Bennett relation~(\ref{bennett}) by means of the statistical concept of maximum likelihood. Here, this approach is based on a transformation of the Crooks relation~(\ref{crooks}) into an expression for the probability $P(f|W)$ with which %the work $w$  results from a
the forward protocol is drawn from an ensemble of equally many realizations of both protocols under the condition that the work performed on the system is $w=W$, for the forward, and $w=-W$ for the backward protocol. 
This conditional probability takes the form of a Fermi function, reading~\cite{shirts}
\begin{equation}
P(f|W) =\frac{1}{1+\exp\left [- \beta(W-\Delta F) \right ]}\:.
\label{Pfw}
\end{equation}
%\begin{equation}
%P(f|W) =\frac{1}{1+\exp \left[- \beta(W}-\Delta F) \right]}\:.
%\label{Pfw}
%\end{equation}
The complementary probability $P(b|W)=1-P(f|W)$ for finding $W$ in a realization of the backward process then becomes
\begin{equation}
P(b|W) =\frac{1}{1+\exp\left [ \beta(W-\Delta F) \right ]}\:.
\label{Pbw}
\end{equation}
%\begin{equation}
%P(b|W) =\frac{1}{1+\exp\left[ \beta(W-\Delta F) \right]}\.
%\label{Pbw}
%\end{equation}
According to the maximum likelihood method, the most likely value of the free energy change compatible with the work values of M realizations of the forward and equally many realizations of the backward protocol
maximizes the likelihood defined as the joint probability $\ell(\Delta F) = \prod_j^M P(f|w_{j,f})P(b|-w_{j,b})$  evaluated at the actual outcomes $w_{j,\alpha}$, $j=1..M$, of the forward, $\alpha=f$, and the backward, $\alpha = b$ protocols.
This leads to
\begin{equation}\label{bennettfinite}
\frac{1}{M}\sum_{j=1}^{M} \frac{1}{1+e^{\beta(w_{j,f}-\Delta F)}}=
\frac{1}{M}\sum_{j=1}^{M} \frac{1}{1+e^{\beta(w_{j,b}+\Delta F)}}.
\end{equation}
presenting a non-linear equation in $\Delta F$ which, for given data $w_{j,f}$ and $w_{j,b}$,
has a uniquely defined solution. It can numerically be solved by means of the Newton algorithm.
In the continuum limit of $M \to \infty$ this equation can be written as
\begin{equation}\label{contb}
\int_{-\infty}^{\infty} dw \frac{p_{f}(w)}{1+e^{\beta(w-\Delta F)}}=\int_{-\infty}^{\infty} dw\frac{p_{b}(w)}{1+ e^{\beta(w+\Delta F)}},
\end{equation}
yielding Eq.~(\ref{bennett}) upon a change of the variable $w\rightarrow - w$ on the right hand side.

In the limiting cases of slow and rapid protocols, Eqs.~(\ref{bennettfinite}) and (\ref{contb}) have simple solutions. %in limiting cases.
When the work protocol is performed quasi-statically so that the associated dissipation is small,
the majority of work values are localized near the free energy difference such that $|w-\Delta F| \ll k_{B}T$ for the forward, and $|w+\Delta F| \ll k_{B}T$ for the backward protocol. The expansions of the exponential factors in Eq.~(\ref{contb}) in terms of their small arguments lead to
\begin{equation}
\int_{-\infty}^{\infty}dw (w-\Delta F)p_{f}(w)\approx \int_{-\infty}^{\infty}
dw (\Delta F-w) p_{b}(-w)
\end{equation}
yielding for the free energy change
\begin{equation}
\langle w\rangle_{f}-\langle w\rangle_{b}\approx 2\Delta F.
\label{BG}
\end{equation}
For a slow work protocol with small dissipations, the Bennett estimate of the free energy difference is given by
the difference of averaged works of forward and backward protocols.

On the other hand, for a fast protocol generating large dissipation, the overwhelming majority of forward data satisfies  $w_{j,f}-\Delta F \gg k_B T$ and accordingly  $w_{j,b}+\Delta F \gg k_BT$ for the backward data. Then, we can expand the Fermi-functions in  Eq.~(\ref{bennettfinite}) giving in leading order
\begin{equation}
\frac{e^{2\beta\Delta F}}{M} \sum_{j=1}^{M}e^{-\beta w_{j,f}} \approx  \frac{1}{M}\sum_{j=1}^{M}e^{-\beta w_{j,b}} .
\end{equation}
This yields a central result of this work
\begin{equation}\label{half}
2\Delta F\approx \Delta F_{f}-\Delta F_{b} \equiv 2\Delta F_{H},
\end{equation}
where the Jazynski estimates $\Delta F_{\alpha}$ are determined from $M$  realizations
of the forward ($\alpha= f$) and the backward ($\alpha =b$) protocols as,
\begin{equation}\label{sb}
\beta \Delta F_{\alpha}\equiv -\ln\left[\frac{1}{M}\sum_{j=1}^{M}e^{-\beta w_{j,\alpha}}\right]\:.
\end{equation}
The Bennett estimate $\Delta F$ in the large dissipation regime is given by half of the difference of the forward and the backward Jarzynski estimates,
$\Delta F_{f}-\Delta F_{b}$, which we denote as $\Delta F_{H}$.

Due to the finite sampling of
work values, the estimates $\Delta F_{\alpha}$ are still random and hence can differ from experiment to experiment.
In order to characterize the statistics of the Jarzynski estimator for a finite number $M$ of work data, one considers $m$ repetitions of $M$ work measurements.
%In this respect, consider $m$ different experiments and for each experiment we sample $M$ work values.
The totality of work data then consists of
$\{[w^{(1)}_{j}],[w^{(2)}_{j}],\cdots , [w^{(m)}_{j}]\}$ where $[w^{(\ell)}_{j}]$ is the data set from the $\ell$-th experiment:
$[w^{(\ell)}_{j}]\equiv \{w^{(\ell)}_{1},w^{(\ell)}_{2},\cdots w^{(\ell)}_{M}\} $.
Based on the resulting set of the Jarzynski estimates for $M$ data according to Eq.~(\ref{sb}) the statistics of these estimates can be analyzed. In particular, one obtains % Then one can determine
a block averaged estimate of the free energy difference from $m$ experiments reading
\begin{equation}\label{jeba}
\beta \overline{\Delta F_{\alpha}} =- \lim_{m\rightarrow \infty}\frac{1}{m}\sum_{\ell =1}^{m}
 \ln\left[\frac{1}{M}\sum_{j=1}^{M}e^{-\beta w^{(\ell)}_{j,\alpha}}\right].
 \end{equation}
 It was proven that the block averaged Jarzynski estimator, $\overline{\Delta F_{\alpha}}$, with a finite $M$ is bounded from below by the true value of $\Delta F$ and from above by the average work $\langle w \rangle_{\alpha}$, i.e., it lies in the range~\cite{zucker2,jarzynski3}
 \begin{equation}\label{mono}
 \Delta F \leq \overline{\Delta F_{\alpha}} \leq \langle w\rangle_{\alpha}.
\end{equation}
Unfortunately, neither a lower nor an upper bound of the Bennett estimator is known.
The underlying difficulty %is suggested by
results from the structure of the Eq.~(\ref{half}), which is based on %where
the {\it difference} between the forward and backward Jarzynski estimates.
Therefore the inequalities~(\ref{mono}) valid for the individual terms do not translate
% gives
to the Bennett estimation in the large dissipation limit.

\section{Gaussian chain}
We illustrate the aforementioned features by considering a one-dimensional Gaussian chain of $(N+1)$ beads connected by
harmonic springs. Additionally, the beads experience strong friction and fluctuating forces stemming from a heat bath at temperature $T$.
The potential energy of the system is given by
\begin{equation}\label{harmonic}
U(\{x_{i}\})=\frac{k}{2}\sum_{i=1}^{N}(x_{i}-x_{i-1})^{2},
\end{equation}
where $x_{i}$ denotes the  position of the $i$th bead; the first bead with $i=0$ is fixed at the origin~($x_{0}=0$).
A simple way to perform work on the system is to pull the last bead~($i=N$) with a constant speed. The forward protocol acts
during a time interval $t\in (0,t_{f})$, starting from the thermal equilibrium state of the chain with its end at the origin, $x_{N}(0)=0$, and proceeds by increasing $x_{N}(t)$ linearly in time as $x_{N}(t)=vt$ until the chain end reaches the designated position $x_{N}(t_{f})=x_{d}$. For the backward protocol, the chain is initially in the thermal equilibrium while its end  is fixed at $x_{d}$ and the work is done by changing the chain end position as $x(t)=x_{d}-vt$.
The Jarzynski work for the forward protocol is given by
\begin{equation}\label{work}
w=v \int_{0}^{t_{f}}dt\frac{\partial U(\{x_{i}\})}{\partial x_{N}} = v k \int_0^{t_f} dt [x_N(t) - x_{N-1}(t) ],
\end{equation}
and the same expression with $v \rightarrow -v $ gives the work for the backward protocol.
The overdamped motion of the beads~($i=1,2,\cdots,N-1$) with the friction constant $\gamma$ can be
described by a position Langevin equation:
\begin{equation}
\gamma \frac{d x_{i}}{d t}=-k(x_{i+1}+x_{i-1}-2x_{i})+\xi_{i}(t),
\label{legc}
\end{equation}
where the Gaussian white noises  $\xi_i(t)$ model thermal random forces exerted on the chain.
Accordingly, they satisfy $\langle \xi_{i}(t)\rangle =0$ and $\langle \xi_{i}(t)\xi_{j}(t')\rangle = 2\gamma k_{B}T\delta_{i,j}\delta(t-t')$.
%This corresponds to
Eq.~(\ref{legc}) describes the Rouse model of a polymer in a viscous fluid where the hydrodynamic interactions are neglected.

%%%%%%%%%%%%%%%%%%%%%%%%%%%%%%%%%%%%%%%%%%%%%%%%%%%%%%%%%%%%%%%%%%%%%%%%%%%%%%%%%%%%%%%%%%%%%%%
\begin{figure}[t]
%\resizebox{7cm}{!}{
\includegraphics[width=.9\columnwidth]{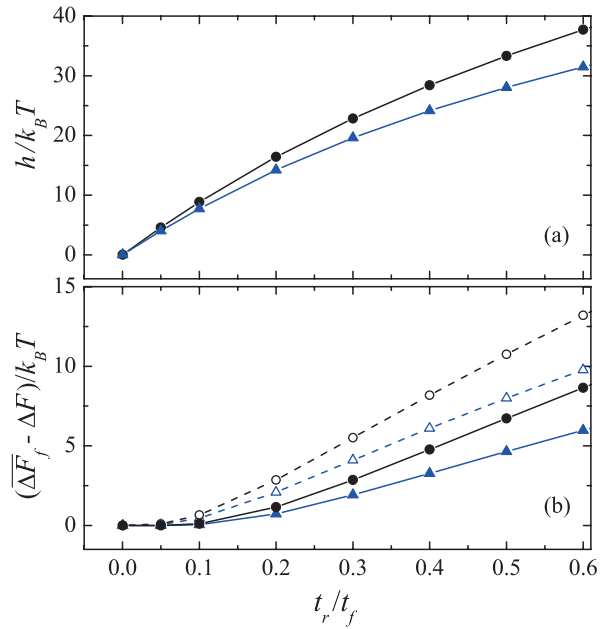}
%}
 \caption{
(Color online)~(a) The work hysteresis $h$ as defined in Eq.~(\ref{hysteresis}) is displayed as a function of the time scale ratio $t_{r}/t_{f}$ representing the pulling speed, for $\Delta F =15 k_B T$, and for different lengths of the chain, $N=40$~($\bullet$) and $N=10$ ($\blacktriangle$) in the regime of slow pulling speeds. Note that $h/k_b T$ agrees with the work variance of the Gaussian chain. For vanishing pulling speed ($t_r/t_f \to 0$) the variance vanishes.
At higher pulling speeds, the chain-length dependence is more pronounced. The panel (b) shows the deviations of the block averaged Jarzynski estimator as defined in Eq.~(\ref{jeba}) from the exact value of the free energy change as a function of pulling speed.
The data were sampled from the exactly known Gaussian work distributions given by the Eqs. (\ref{gauss}) and (\ref{avesig}),
rather than by simulations of the Langevin equations~(\ref{legc}).
%instead of being obtained from Langevin dynamics simulations using Eq.~(\ref{legc}).
The sampling size was chosen as $M=10^{4}$ ($\circ$ for $N=40$ and $\bigtriangleup$ for $N=10$) and $M=10^{6}$ ($\bullet$ for $N=40$ and $\blacktriangle$ for $N=10$).
In both cases the number of $m=3\times 10^{2}$ blocks led to well converged block averages. % with negligible  statistical errors.
As the pulling speed increases, the error due to the finite sampling size becomes significant. Only the protocol taking much longer than the relaxation time of the system~($t_{r}/t_{f} \lesssim {\cal O}(10^{-2})$) yields an error less than $k_{B}T$. In both panels, the lines connecting the symbols are guides to the eye.
}
\label{fig3}
\end{figure}
%%%%%%%%%%%%%%%%%%%%%%%%%%%%%%%%%%%%%%%%%%%%%%%%%%%%%%%%%%%%%%%%%%%%%%%%

%%%%%%%%%%%%%%%%%%%%%%%%%%%%%%%%%%%%%%%%%%%%%%%%%%%%%%%%%%%%%%%%%%%%%%%%%%%%%%%%%%%%%%%%%%%%%%%
\begin{figure}[t]
%\resizebox{7cm}{!}{
\includegraphics[width=.9\columnwidth]{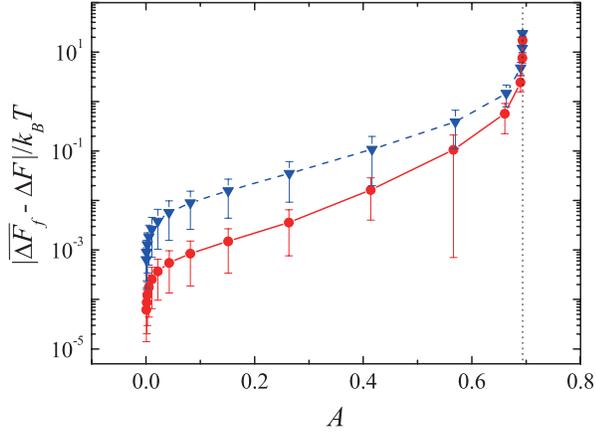}
%}
 \caption{
(Color online)~The Jarzynski bias as a function of the time asymmetry for $N=40$ for pulling protocols with $\Delta F = 15 k_B T$. The duration of time in which the final end to end distance was reached was chosen as $t_{f}= 2^{-n} 10^3 t_{r}$ with $n$ varying from $n=0$ up to $n=15$ for different block sizes, $M=10^{4}$ ($\blacktriangledown$) and $M=10^{6}$~($\bullet$).  Averages are performed over $m= 3\times10^{2}$ blocks. Irrespective of the block size $M$, the bias remains within $k_{B}T$ when the time asymmetry is less than $\ln 2$~(represented by the vertical dotted line), and abruptly rises in the limit of maximal time asymmetry, $A\to \ln 2$. The broken and solid lines connecting equal symbols serve as guides to the eye. The vertical bars at the symbols indicate the according variances of the Jarzynski estimator.
}
\label{fig4}
\end{figure}
%%%%%%%%%%%%%%%%%%%%%%%%%%%%%%%%%%%%%%%%%%%%%%%%%%%%%%%%%%%%%%%%%%%%%%%%
This model is exactly solvable as shown by Dhar~\cite{dhar}. The free energy difference associated with
stretching in the forward process is given by
\begin{equation}\label{deltaF}
\Delta F = \frac{k}{2} \frac{x_{d}^{2}}{N}.
\end{equation}
and accordingly $-\Delta F$ for the backward process.
The probability distribution of the work is of Gaussian form as given by the
Eq.~(\ref{gauss}), with the forward work average and the variance~\cite{dhar},
\begin{eqnarray}\label{avesig}
\langle w\rangle_{f}&=&\Delta F + \frac{\beta}{2}\sigma^{2}, \\ \nonumber
\sigma^{2}&=&2\beta^{-1}\gamma v x_{d}[{\mathcal L}^{-2}+{\mathcal L}^{-3}(e^{-{\mathcal L}\tau}-1)/\tau]_{N-1,N-1},
\end{eqnarray}
where ${-\mathcal L}$ is the lattice Laplacian, ${\mathcal L}_{ij}=2\delta_{i,j}-\delta_{i,j+1}-\delta_{i,j-1}$, and
$\tau=(k/\gamma)t_{f}$.
For the backward protocol, the PDF is also Gaussian with the same variance, $\sigma$ as given in Eq.~(\ref{avesig}), and the work average
\begin{equation}\label{workb}
\langle w\rangle_{b}=-\Delta F +\frac{\beta}{2}\sigma^{2},
\end{equation}
indicating that the free energy difference for the Gaussian distribution is determined by the difference of the forward and backward work averages:
$2\Delta F = \langle w\rangle_{f}-\langle w\rangle_{b}$. On the other hand, the hysteresis, defined by Eq.~(\ref{hysteresis})
becomes $h= \beta\sigma^{2}/2$. %{\color{red} For $\tau \ll 1$}
For a given final extension the pulling
speed is a crucial factor in determining the value of $\sigma$, as it appears in the prefactor. According to Eq.~(\ref{avesig}), for $\tau \to 0$, i.e. for a sudden change of the end-to-end distance, %chain length,
the variance approaches  a finite value which is almost independent of the chain length~\cite{dhar}. With increasing duration of the protocol the work variance becomes dependent on the chain length, indicating a collective response of the chain.  In the limit of infinitely slow processes the variance asymptotically vanishes as $1/\tau$, as one would expect for  an isothermal quasi-static protocol.  %{\color{red} ({\it Why $\tau \ll 1$? The prefactor does always contain  $v$}.)} {\color{red} For $\tau \gg 1$} %very slow pulling speeds,
%the variance linearly increases in $v$ as $\sigma \sim v N$, where the chain length comes into play, indicating a collective response of the total chain.
%For fast pulling speeds, mainly the chain end responds to the perturbation and yields a variance that is almost independent of $N$~\cite{dhar}.
``Slow'' and ``fast'' can be quantified relative to
characteristic time scales of the system, for example, to its relaxation times. For  Gaussian chains considered here the relaxation time is given by $t_{r}=\gamma/(k\lambda_{m})$ with
$\lambda_{m}$ being the minimum eigenvalue of the negative lattice Laplacian ${\mathcal L}$:
$\lambda_{m}=2-2\cos [\pi/ N]$.
For $N \gg 1$, the relaxation time increases quadratically with the chain length.
Since we are interested in the efficiency of different estimators of free energy change we compared these for two different chain lengths and different protocols, in all cases leading to the same change of the free energy. According to Eq.~(\ref{deltaF}) the final extension then depends on the chain length as $x_d \propto \sqrt{N}$ for large $N$.
The ratio of the two time scales, $t_{r}/t_{f}=v/(\lambda_m \sqrt{2 N \Delta F/k}) $ then quantifies the progress rate of the protocol: In particular, for $t_{r}/t_{f} \ll 1$, the protocol approaches a reversible processes with vanishing dissipated work.

%%%%%%%%%%%%%%%%%%%%%%%%%%%%%%%%%%%%%%%%%%%%%%%%%%%%%%%%%%%%%%%%%%%%%%%%%%%%%%%%%%%%%%%%%%%%%%%
\begin{figure}[t]
%\resizebox{7cm}{!}{
\includegraphics[width=.9\columnwidth]{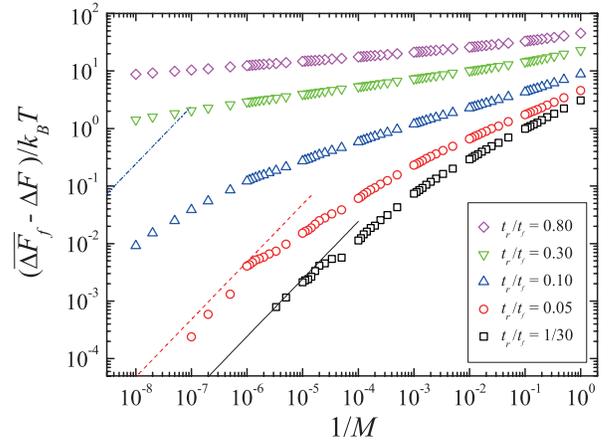}
%}
 \caption{
(Color online)~The finite sampling behavior of the Jarzynski estimator for $\Delta F = 15k_{B}T$, and various progress rates of the protocol, $t_{r}/t_{f}$, as a function of the sampling size $M$.
Averages are performed over $m= 10^{4}$ blocks.
The estimated free energy difference shows the monotonic behavior %following from
predicted by Eq.~(\ref{jeba}), and is larger than the true value for any finite $M$ in agreement with Eq.~(\ref{mono}). The lines represent the large-$M$ asymptotic behavior, Eq.~(\ref{asymp}). Only the protocols with slow pulling speeds for $t_{r}/t_{f} =1/30$ and $t_{r}/t_{f} = 0.05$ display this asymptotic behavior. In order to see it for faster pulling speeds one has to go to even larger sample sizes.
}
\label{fig5}
\end{figure}
%%%%%%%%%%%%%%%%%%%%%%%%%%%%%%%%%%%%%%%%%%%%%%%%%%%%%%%%%%%%%%%%%%%%%%%%

For given $\Delta F$ and $t_{r}/t_{f}$, the exact value of the variance $\sigma$
can be found using Eqs.~(\ref{deltaF})-(\ref{workb}), which completely characterize the Gaussian work PDF given by Eq.~(\ref{gauss})~\cite{dhar}.
This Gaussian distribution indicates the ideal work PDF which one would obtain from an infinite number of measurements,
leading to unbiased results of the estimators.
In order to address the statistical bias for a finite number of data,
we investigated the dependence of various free energy estimators on $t_{r}/t_{f}$ and on the sample size by drawing data
from the exactly known Gaussian work distribution given by the Eqs.~(\ref{gauss}): In this way we could avoid time-consuming simulations of the Langevin equations~(\ref{legc}) and yet obtain large amounts of data with low computational effort.
%we did not perform the Langevin dynamics simulations using Eq.~(\ref{legc}) to evaluate the work for a given trajectory.
%Instead the work values are sampled from a Gaussian work distribution~\cite{bias1}.}
%In this way large amounts of data could be obtained in a much shorter time than the direct simulations of the Langevin equation would have taken.

First, in Fig.~3(a), we present results for the hysteresis $h$ as a function of $t_{r}/t_{f}$.
From the value of $h$ at a given $t_{r}/t_{f}$, one can estimate the required number of measurements for a reasonable estimate of $\Delta F$ by the Jarzynski equality according to $M_c \gtrsim e^{\beta h}$. For example, for $t_{r}/t_{f} =0.1$, $M_{c}\sim e^{10}\sim 10^{4}$. If the chain is pulled faster, $M_{c}$ may become enormously large so that the free energy
estimation from a finite number of measurements becomes totally unreliable.
We used the known work PDF of the Gaussian chain as a test case and determined block averages over $m=3\times10^2$ blocks of different sizes.
Figure 3(b) shows the errors in the free energy estimation from the forward work measurement,
%{\cblu
$(\overline{\Delta F_{f}}-\Delta F)/k_B T$, where $\overline{\Delta F_{f}}$ is obtained via Eq.~(\ref{jeba}) for two different sample sizes with $M=10^{4}$ and $M=10^{6}$. With $t_{r}/t_{f}$ the dissipated work increases giving rise to increasing deviations of the estimated free energy change from its true value.

With increasing time asymmetry the total dissipated work grows. For the Jarzynski estimator this growth leads to a systematic increase of its bias relative to the exact value as displayed in Fig.~4. In particular, the bias diverges if the limiting value $\ln 2$ of the time asymmetry is approached. This divergence persists for any finite number of data in accordance with the exponential scaling of the required number of data, $M_c$,
with the hysteresis, $M_c \propto e^h$.
The bias of the Jarzynski estimator is presented
in Fig.~5 as a function of the block size $M$ for different pulling speeds.
It asymptotically approaches zero at different rates depending on the pulling speed for large values of $M$.
According to Ref.~\cite{zucker1,zucker2} the asymptotic approach is given by a power law reading:
\begin{equation}\label{asymp}
(\overline{\Delta F_{\alpha}}-\Delta F)/k_BT = (e^{\beta^2 \sigma^{2}}-1)/(2M) + {\cal O}(M^{-2}).
\end{equation}
This behavior though can only be identified for the two cases with slowest pulling speeds $t_{r}/t_{f}=1/30$ and $t_{r}/t_{f}=0.05$ (See the straight lines in Fig.~5).
For the other, faster protocols, the sample sizes examined here are insufficient to enter
the asymptotic regime being governed by the central limit theorem.
Instead, we observe algebraic decay behavior as $1/M^{\alpha}$ with $\alpha < 1$. For the sample sizes studied here,
the decay exponent $\alpha$ becomes very small with increasing pulling speed, signaling a bad convergence of the
finite sample average for fast protocols.

%%%%%%%%%%%%%%%%%%%%%%%%%%%%%%%%%%%%%%%%%%%%%%%%%%%%%%%%%%%%%%%%%%%%%%%%%%%%%%%%%%%%%%%%%%%%%%%
\begin{figure}[t]
%\resizebox{7cm}{!}{
\includegraphics[width=.9\columnwidth]{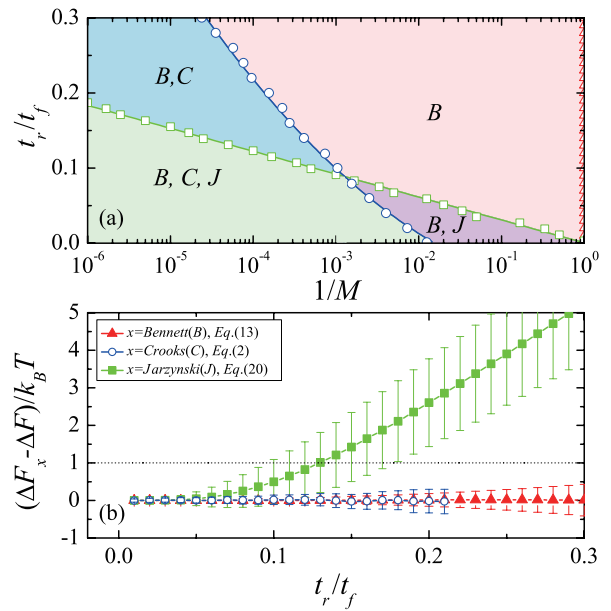}
%}
 \caption{
(Color online)~(a) Efficiency diagram of the three methods for $N=40$ and $\Delta F =15k_{B}T$.
Each sample of size $M$ was independently generated $m=10^4$ times.
The symbols J, C, and B are assigned to those regions in the parameter plane spanned by the inverse box size and $t_{r}/t_{f}$, where the Jarzynski,  Crooks, and Bennett methods, respectively, give an estimate of the free energy difference with an error being less than $k_{B}T$.
Along the line connecting the squares, the expected error of the Jarzynski estimator equals $k_BT$, i.e.,
$\overline{\Delta F_{f}}-\Delta F = k_{B}T$. Below this line, for smaller pulling speeds, the error is smaller. The line connecting the circles indicates the corresponding curve for the Crooks estimator.
(b) Estimated bias of three estimators of free energy difference as a function of $t_{r}/t_{f}$ with $M=10^{4}$
and $m=3\times10^2$. The Jarzynski error becomes larger than $k_{B}T$ as soon as $t_{r}/t_{f}\gtrsim 0.1$. The Crooks crossing criterion
gives the correct value of $\Delta F$ up to $t_{r}/t_{f} \approx 0.2$, where the overlap between
the forward and the backward PDF disappears so that the Crooks criterion only delimits the range within which the free energy difference is located. Apart from an increase of the expected sampling error,
Bennett's method according to Eq. (\ref{bennettfinite}) is insensitive to the pulling speed, yielding precise
estimates of $\Delta F$.
The error bars indicate the magnitude of the variance of the free energy estimators based on $m=3\times10^2$ blocks of size $M=10^4$.
The variance of the Jarzynski estimator increases rapidly with increasing speed, while the variances of the Crooks and the Bennett estimators grow much slower.
}
\label{fig6}
\end{figure}
%%%%%%%%%%%%%%%%%%%%%%%%%%%%%%%%%%%%%%%%%%%%%%%%%%%%%%%%%%%%%%%%%%%%%%%%

Our next goal is to systematically compare the biases of three estimators, the Jarzynski equality, Crooks' crossing criterion,
and the Bennett method,
for the Gaussian work PDF as exemplified by the pulling protocol of the Gaussian chain.
In Fig~6(a), we display the absolute magnitude of the biases of the three estimators as functions of $t_{r}/t_{f}$ and sampling size $M$.
As expected, the Bennett method is always superior to the other estimators.
The symbols J, C and B in Fig. 6~(a) indicate those parameter regions within which the Jarzynski, Crooks and Bennett estimators, respectively, deviate by less than the thermal energy $k_B T$ from the true free energy difference.
The Jarzynski estimator soon becomes unreliable with increasing pulling speed, while the Crooks estimator yields better values for larger pulling speeds provided that it is based on a sufficiently large number of data. It is interesting to note that for small $M$ and very slow pulling speeds the Jarzynski estimator performs better than the Crooks estimator. %at small $M$.
To better understand this observation, note that already for $M=1$, the block average ~(\ref{avesig}) yields
$\overline{\Delta F_{f}}=\langle w\rangle_{f}$ which is close to the correct value for weakly dissipative protocols, according to  Eq.~(\ref{jeba}).
On the other hand, for the Crook's crossing criterion to properly work, the tails of the forward and backward PDFs need to be sampled with sufficient accuracy,
necessitating a sample of reasonable size.
The Bennett estimator for this model with a Gaussian work PDF is found to be free of any bias in the whole investigated parameter region. Yet with increasing pulling speed the variance of the Bennett estimator increases as displayed in panel (b) of Fig.~6 for a fixed $M$.
This increase though is less pronounced than those of the respective variances of the Crooks and the Jarzynski estimators.
%{\it whether the variance of the Crooks is less than that of the Bennett is not clearly visible; we can
%simply quote the variance of the Jarzynski only.}

%%%%%%%%%%%%%%%%%%%%%%%%%%%%%%%%%%%%%%%%%%%%%%%%%%%%%%%%%%%%%%%%%%%%%%%%%%%%%%%%%%%%%%%%%%%%%%%

\section{Chain with a hairpin}

%%%%%%%%%%%%%%%%%%%%%%%%%%%%%%%%%%%%%%%%%%%%%%%%%%%%%%%%%%%%%%%%%%%%%%%%%%%%%%%%%%%%%%%%%%%%%%%
\begin{figure}[t]
%\resizebox{7cm}{!}{
\includegraphics[width=.9\columnwidth]{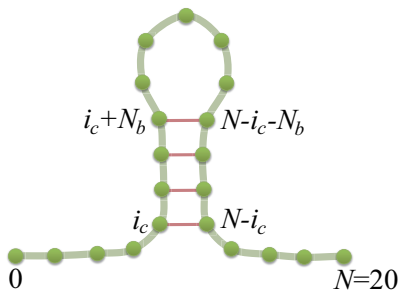}
%}
 \caption{
(Color online)~A schematic picture of a hairpin chain with $N=20$. The pairing interactions are present
between the $i$-th and $N-i$-th monomers where $i=i_c, i_c+1, \cdots, i_c+N_b$ with $N_b+1$ denoting the number of pairs.
In this work, we considered a hairpin chain with $N_b=3$.
}
\label{fighp}
\end{figure}
%%%%%%%%%%%%%%%%%%%%%%%%%%%%%%%%%%%%%%%%%%%%%%%%%%%%%%%%%%%%%%%%%%%%%%%%
%%%%%%%%%%%%%%%%%%%%%%%%%%%%%%%%%%%%%%%%%%%%%%%%%%%%%%%%%%%%%%%%%%%%%%%%%%
\begin{figure}[t]
%\resizebox{7cm}{!}{
\includegraphics[width=.9\columnwidth]{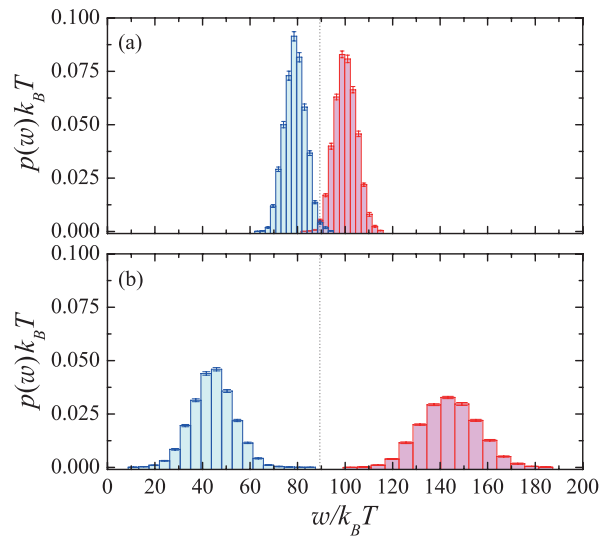}
 \caption{
(Color online)~Histograms of the work performed on a hairpin chain as depicted in Fig.~7.
%In the inset of panel (a) the configuration of the chain is displayed schematically.
The chain parameters are $N=20$, $\tilde{k}=30$ and $\tilde{\epsilon}=20$; the maximal end to end distance of the chain reached at the end of the forward protocol is $x_d=25 a$; for further details
see the text below Eq.~(\ref{lehp}). Each histogram was obtained as an average over
$m=10^2$ independently generated histograms based on $M=10^4$ simulations of the Langevin equation (\ref{lehp}). The variances of the such estimated probabilities indicated by error bars are relatively small.
On the right hand side the histograms of the forward protocols and on the left hand side those for negative work of the backward protocols are depicted.
Panel (a) displays the case when the protocol lasted $t_{f}=900 t_{r}^{3D}$ representing a relatively slow pulling speed for which the forward and backward work histograms overlap. Note that here $t_{r}^{3D}$ is the relaxation time of a three-dimensional
Gaussian chain, $t_{r}^{3D}=t_{r}/3=\gamma/(3k\lambda_{m})$. Consequently a reliable estimate of $\Delta F = 89.4k_{B}T$ indicated by the vertical dotted line was obtained from the Crooks crossing criterion. Panel (b) gives the work distributions for a five times faster pulling speed~($t_{f}=180 t_{r}^{3D}$). At this higher speed the histograms become broader and more dissimilar and no longer overlap with each other. }
\end{figure}
%%%%%%%%%%%%%%%%%%%%%%%%%%%%%%%%%%%%%%%%%%%%%%%%%%%%%%%%%%%%%%%%%%%%%%%%
A particular feature of a Gaussian work distribution given by Eq.~(\ref{gauss}), such as the one for the
pulling process of a Gaussian chain, is  the mirror-symmetry of the forward and backward distributions  with respect to $\Delta F$ relating $p_f(w)$ and $p_b(-w)$.
As a consequence, the dissipated works are equal for the forward and the backward processes,
and,  moreover, the Bennett estimator is unbiased (See Fig.~6(b)).
In order to demonstrate the dependence of the Bennett estimator on the difference of the forward and backward dissipated works,
we thus need to consider an asymmetric work PDF.
As shown in the first experimental realization of the Jarzynski equality~\cite{sm1},
pulling a hairpin molecule typically exhibits pronounced asymmetric work PDFs, depending upon the direction of the protocol.
Therefore, we consider a chain in three dimensions consisting of $N$ monomers with Hookean bonds along the chain and,  additionally, pair-specific interactions leading to a hairpin structure in the mechanical equilibrium state, as depicted in Fig~7.
The potential energy of this chain is given by %{\bf[the second summation is rewritten, not sure if it is clearer]}
\begin{equation}\label{hairpin}
\begin{split}
U(\{\mathbf{r}_{i}\})&=\frac{k}{2}\sum_{i=1}^{N}(r_{i, i-1}-a)^{2}\\
&\quad +\epsilon\sum_{i=i_{c}}^{i_{c}+N_{b}}
\left[\left(\frac{a}{r_{i,N-i}}\right)^{12} -\left(\frac{a}{r_{i,N-i}}\right)^{6}\right],
\end{split}
\end{equation}
where ${\mathbf r}_{i}=(x_{i},y_{i},z_{i})$ denotes the position of the $i$-th monomer, and $r_{i,j}= |{\mathbf r}_{i}-{\mathbf r}_{j}|$ the distance between two monomers. The first sum represents the contribution of the Hookean springs %energy between the
connecting neighboring monomers along the chain. %contour,
Here $a$ is the equilibrium bond length and $k$ the spring-constant.
The second sum describes the interaction between pairs of monomers $i$ and $N-i$, given by the Lennard-Jones potential.
The summation index runs over the monomers constituting the pairs as depicted
%where the summation runs over the one of the monomers in pair, and $N_{b}+1$ denotes the total number of pairs~
in Fig~7. The total number of pairs is $N_b+1$. % indicates the total number of pairs.}
As for the Gaussian case, additionally, strong friction and random forces act on the beads of the chain resulting in an overdamped dynamics which is governed by the %position
Langevin equation
\begin{equation}
\gamma \frac{d {\mathbf r}_{i}}{d t}=-\nabla_{i}U(\{\mathbf{r}_{i}\})+{\boldsymbol \xi}_{i}(t),
\label{lehp}
\end{equation}
where $\gamma$ denotes the friction constant experienced by a monomer,
$\nabla_{i}$ the three dimensional gradient with respect to the position of the $i$th monomer, and
${\boldsymbol \xi}_{i}(t)=(\xi_{i,x}(t),\xi_{i,y}(t),\xi_{i,z}(t))$ thermal Gaussian white noise forces
satisfying
$\langle \xi_{i,\ell}(t)\rangle =0$ and
$\langle \xi_{i,\ell}(t)\xi_{j,m}(t')\rangle = 2 \gamma k_{B}T\delta_{i,j}\delta_{\ell,m}\delta(t-t')$.
Similarly as for the Gaussian chain, the one end of a hairpin chain is fixed at ${\mathbf r}_0=0$ and the other end ${\mathbf r}_N$ is pulled with constant speed $v$ in the $x$-direction.
As for the Gaussian chain, the work performed in this way is given by
$v \int_0^{t_f} dt \partial U(\{{\mathbf r}_i\})/\partial x_N$ with the hairpin potential, Eq.~(\ref{hairpin}).

Unlike the case of pulling a Gaussian chain, for a pulled hairpin chain, we do not know the exact work distribution
from which we could sample the data. In order to study the effect of an asymmetric work PDF on the biases,
we had to %explicitly
perform direct simulations of the Langevin dynamics.
For the numerical simulations, we rescaled all lengths by the bond length according to ${\tilde{\mathbf r}} ={\mathbf r}/a$
and discretized the Langevin equation with a time step $\Delta$.
The iterative Langevin equation then reads in terms of the discrete time variable $n=t/\Delta$ as
\begin{equation}
\tilde{\mathbf r}_i (n+1) = \tilde{\mathbf r}_i (n) -\tilde{\mu} \tilde{\nabla}_i \tilde{U}(\{{\mathbf r}(n)\})+\tilde{{\boldsymbol \xi}}_i(n),
\label{dle}
\end{equation}
where $\tilde{U}=U/k_BT$ is the rescaled potential, and $\tilde{\mu}$ =  $\Delta/t_a$ denotes the rescaled mobility with $t_a=\gamma a^2/(k_B T)$ being a characteristic time-scale of the chain.
The variances of the dimensionless Gaussian random forces $\tilde{\boldsymbol \xi}_i(n)$ are determined by
$\langle \tilde{\xi}_{i,k}(m) \tilde{\xi}_{j,\ell}(n)\rangle = 2 \tilde{\mu} \delta_{m,n} \delta_{k,\ell} \delta_{i,j}$.
The dimensionless potential $\tilde{U}$ is determined by the dimensionless spring constant $\tilde{k}=k a^2/k_BT$ and the dimensionless energy parameter of the Lennard-Jones potential
$\tilde{\epsilon}=\epsilon/k_BT$. In the simulations these parameters were set as $\tilde{k}=30$ and $\tilde{\epsilon}=20$.
For the efficiency as well as the numerical convergence of the simulations,
the rescaled mobility was chosen as $\tilde{\mu} = 0.0001$.
We considered a hairpin-like molecule with N=20 monomers having pairing potentials between the monomer pairs $(4,16)$ up to $(7,13)$.

Before the forward protocol was started, the chain had been prepared into a thermal equilibrium state having the form of a hairpin with constrained positions of the first and last monomers, ${\mathbf r}_0=0$, ${\mathbf r}_N(0)=a\hat{{\mathbf x}}$, respectively, where $\hat{{\mathbf x}}$ denotes the unit vector in $x$-direction. This thermal equilibrium state was established
by simulating the Langevin equation~(\ref{dle}) for a sufficiently large time with clamped end positions.
Upon equilibration, the last monomer was pulled at a constant speed $v$ in the $x$-direction
until it had reached the distance $x_d=|{\mathbf r}_N(t_f)-{\mathbf r}_N(0)|=25a$.  The backward protocol was started with a thermal equilibrium
distribution with the first and
last monomer at the final positions of the forward protocol; then the chain  was compressed by moving the last monomer
at the same absolute velocity $v$ in the $-x$-direction until it had reached the initial position of the forward protocol, ${\mathbf r}_N=a \hat{{\mathbf x}}$.

The histograms displayed in Fig.~8 represent averages of $m=10^2$ raw histograms each based on $M=10^4$ simulations of the Langevin equation for forward and backward protocols at two different pulling speeds. The statistical uncertainty of these averages were estimated from the variance of the distributions of raw histograms and are indicated by error bars.
For the slow protocol displayed in the upper panel, the forward and the backward histograms cross at $\Delta F = 89.4$ indicated by the vertical dotted line, and appears as almost symmetric about the crossing point.
In contrast, for the fast protocol shown in the panel (b), the two histograms do not overlap, and therefore, do not allow to extract $\Delta F$ from the Crooks relation.
Moreover, these histograms are no longer mirror symmetric; the forward histogram is significantly more dispersed than the backward histogram. We also present the biases of the free energy estimators in Fig.~9(a).
As the protocol speeds up, the bias of the Jarzynski estimators (open squares)  becomes more pronounced
and, at the same time, the dissipated work (triangles) grows.
We note that the 1/2-formula (open diamonds) given by Eq.~(\ref{half}) perfectly coincides with the Bennett estimation (filled red triangles). Both methods lead to a significantly smaller bias than the Jarzynski estimate.

Unlike the Gaussian case,
the dissipated work is larger during the forward than during the backward process; see the curve for $\langle w\rangle_{f}-\Delta F$
in comparison with the curve for $\langle w\rangle_{b}+\Delta F$ in Fig.~9(a).
Figure~9(b) presents the bias of the Bennett estimator as a function of the difference between these dissipated works. The monotonic increase of both quantities as functions of the time-ratio $t_r/t_f$ leads to a proportionality
between the bias of the Bennett estimator and the difference between the dissipated works in the forward and backward protocols confirming the previous conjecture.
%shows that the difference of the two values  increases with the protocol speed.
%As conjectured in the previous sections, this clearly indicates that the bias of the Bennett method
%is proportional to the difference of the dissipated forward and backward works.
The influence of the difference between the dissipated forward and backward works on the Bennett estimator can be understood qualitatively, at least in the limit of large dissipation. According to Eq.~(\ref{half}), in this limit, the Bennett estimator can be expressed by half of the difference of the Jarzynski estimators for the forward and the backward protocol. For a Gaussian work distribution the forward and the backward distributions $p_f(w)$ and $p_b(-w)$ are symmetric with respect to $\Delta F$ and consequently, the biases of the two estimators are identical and compensate each other in the 1/2 formula. On the other hand, in all cases with different dissipated works of the forward and the backward processes, the symmetry of the forward and backward distributions is apparently lost and therefore the forward and the backward Jarzynski estimators will have different biases, which then no longer compensate each other in the 1/2 formula.

%This is a natural consequence of the Bennett estimator in the large dissipation limit obtained in Eq.~(\ref{half}), which reads the free energy change as the difference between the Jarzynski estimators from the forward and the backward processes, and hence yields the error dependent on the difference of
%the finite sampling error of $\Delta F_{f}$ and $\Delta F_{b}$ due to the dissipations associated with the forward and the backward process, respectively. When these dissipation are symmetric along the forward and the backward process, the error of $\Delta F_{f}$ and $\Delta F_{b}$ becomes roughly the same, yielding almost unbiased Bennett estimation, which was the case for the Gaussian chain in Sec. IV. On the other hand, the dissipation asymmetry is present in a system such as the hairpin pulling, either of $\Delta F_{f}$ and $\Delta F_{b}$ has larger error than the other and results in the finite error of the Bennett estimation.}

%%%%%%%%%%%%%%%%%%%%%%%%%%%%%%%%%%%%%%%%%%%%%%%%%%%%%%%%%%%%%%%%%%%%%%%%%%%%%%%%%%%%%%%%%%%%%%%
\begin{figure}[t]
%\resizebox{7cm}{!}{
\includegraphics[width=.9\columnwidth]{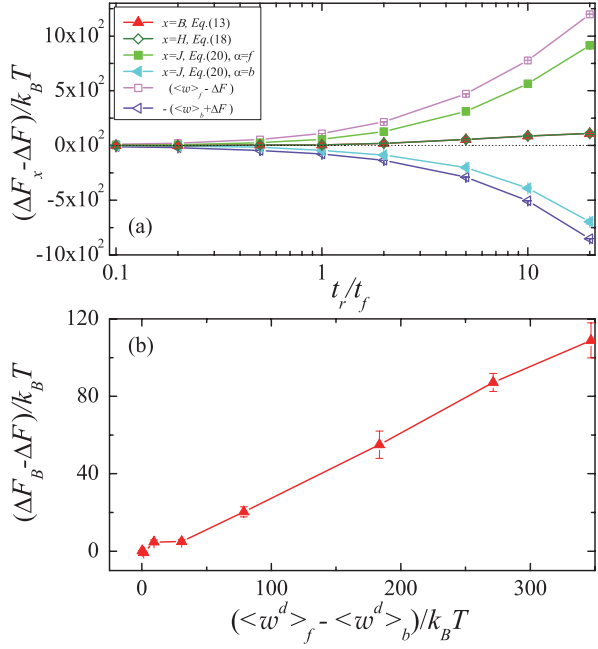}
 \caption{
(Color online)~ (a) The bias of the free energy estimators for %the same
a hairpin chain with parameter values specified in Fig.~8 is depicted as a function of the inverse  duration of the pulling protocol $t_{f}$.
The statistical error bars are smaller than the size of symbols. With increasing
speed, the Jarzynski bias and the dissipated work of the forward and the backward process grow fast. At the same time but at a much slower rate also the difference between the forward and the backward dissipated work
as well as the bias of the Bennett estimator increase.
(b)
The Bennett bias is shown to be proportional to the difference between dissipated works, $\langle w^{d}\rangle_{f}\equiv\langle w\rangle_{f}-\Delta F$ in the forward process and $\langle w^{d}\rangle_{b}\equiv\langle w\rangle_{b}+\Delta F$ in the backward process.
%{\color{red} Comparing the dependence of the dissipated works in the forward and the backward processes symbolized by open squares and open triangles in panel (a), one finds that the difference of the dissipated works increases with the time scale ratio  $t_r/t_f$.  Therefore the bias of the Bennett estimator grows with increasing speed of the protocol.}
%Each points corresponds to the values t_r/t_f in (a) in an increasing order. For the two smallest t_r/t_f values, data points are located near the origin very close to each other.
}
\end{figure}
%%%%%%%%%%%%%%%%%%%%%%%%%%%%%%%%%%%%%%%%%%%%%%%%%%%%%%%%%%%%%%%%%%%%%%%%

\section{summary}
In summary, we discussed the statistical behaviors of the
free energy estimators based on the Jarzynski equality, Crooks' crossing criterion, and Bennett's acceptance ratio method, in relation to the amount of dissipated work
and the time asymmetry. In particular, we
investigated the limiting behaviors of the solutions of Eq.~(\ref{bennett}) and %showed
demonstrated that %when
while both the Jarzynski and the Crooks method are hampered by large %entropy production,
dissipated work, the finite sampling error of Bennett's method is determined by the {\it difference} of %entropy productions
the dissipated works of the forward and the backward processes. As a consequence, it is less severely influenced by the entropy production but rather by the asymmetry between
the forward and the backward process. The examples of a Gaussian and a non-Gaussian chain considered here demonstrate these features.
 The finite sampling error of the Jarzynski estimator rapidly increases with the amount of dissipated work. The Crooks estimator has a binary character: either it yields a reliable solution, as long as the forward and the backward work PDFs overlap, or provides no solution at all. At best an upper and a lower bound can be estimated from the extension of the gap between $p_f(w)$ and $p_b(-w)$.
For Gaussian work PDFs the Bennett method leads to precise estimates over a remarkably wide range of pulling speeds and sampling sizes. For non-Gaussian work PDFs, the bias of the Bennett estimate is still
smaller %compared to
than for other estimators. It, however, increases with increasing differences between the forward and backward dissipated works.

Finally we like to emphasize that
in all cases where the Jarzynski and Crooks estimators fail, the determination of $\Delta F$ can be substantially simplified by means of the 1/2-formula, Eq.~(\ref{half}), expressing the
free energy difference as half of the difference of the forward and backward Jarzynski estimates.

Y.W.K. acknowledges support from Basic Science Research Program through the National Research Foundation of Korea (NRF) funded by the Ministry
of Education, Science and Technology(Grant No. 2010-0025196).
J.Y. acknowledges support from the National Research Foundation of Korea (NRF) grant funded by the Korea government (MEST) (Grant No.  2011-0021296).


\begin{thebibliography}{99}

\bibitem{jarzynski}
C. Jarzynski, Phys. Rev. Lett. {\bf 78}, 2690 (1997).
\bibitem{sm1}
J. Liphardt, S. Dumont, S.~B. Smith, I. Tinoco Jr., C. Bustanamte, Science {\bf 296}, 1832 (2002).
\bibitem{sm2}
N.~C. Harris, Y. Song and C.-H. Kiang, Phys. Rev. Lett. {\bf 99}, 068101 (2007).
\bibitem{pen1}
F. Douarche, S. Ciliberto, A. Petrosyan, and I. Rabbiosi, Euro. Phys. Lett. {\bf 70}, 593 (2005).
\bibitem{pen2}
V. Blickle, T. Speck, L. Helden, U. Seifert, and C. Bechinger, Phys. Rev. Lett. {\bf 96}, 070603 (2006).
\bibitem{jarzynski2}
C. Jarzynski, Phys. Rev. E {\bf 73}, 046105 (2006).
\bibitem{fox}
R.~F. Fox, Proc. Natl. Acad. Sci. USA {\bf 100}, 12537 (2003).
\bibitem{bias1}
J. Gore, F. Ritort, and C. Bustamante, Proc. Natl. Acad. Sci. USA {\bf 100}, %23564
12564 (2003).
\bibitem{bias2}
M. Palassini and F. Ritort, Phys. Rev. Lett. {\bf 107}, 060601 (2011).
\bibitem{zucker1}
D.~M. Zuckerman and T.~B. Woolf, Phys. Rev. Lett. {\bf 89}, 180602 (2002).
\bibitem{zucker2}
D.~M. Zuckerman and T.~B. Woolf, J. Stat. Phys. {\bf 114}, 1303 (2004).
\bibitem{crooks}
G.~E. Crooks, Phys. Rev. E {\bf 60}, 2721 (1999)
%\bibitem{entropy1}
%G.~M. Wang, E.~M. Sevick, E. Mittag, D.~J. Searles, D.~J. Evans, Phys. Rev. Lett. {\bf 89}, 050601 (2002).
\bibitem{entropy2}
J.~M. R Parrondo, C. Van den Broeck and R. Kawai, New J. Phys. {\bf 11}, 073008 (2009).
\bibitem{bennett}
C.~H. Bennett, J. Comput. Phys. {\bf 22}, 245 (1976).
\bibitem{shirts}
M.~R. Shirts, E. Bair, G. Hooker, and V.~S. Pande, Phys. Rev. Lett. {\bf 91}, 140601 (2003).
\bibitem{P}
A. Pohorille, C. Jarzynski, and C. Chipot, J. Phys. Chem. B {\bf 114}, 10235 (2010).
\bibitem{Y}
F. Ytreberg, R.~H. Swendsen, and D.~M. Zuckerman, J. Chem. Phys. {\bf 125}, 184114 (2006).
\bibitem{N}
N. Lu, J.~K. Singh, and D.~A. Kofke, J. Chem. Phys. {\bf 118}, 2977 (2003).
\bibitem{chari}
S.~S.~N. Chari, K.~P.~N. Murthy, and R. Inguva, Phys. Rev. E {\bf 85}, 041117 (2012).
\bibitem{feng}
E.~H. Feng and G.~E. Crooks, Phys. Rev. Lett. {\bf 101}, 090602 (2008).
\bibitem{J}
C. Jarzynski, Annu. Rev. Condens. Matter Phys. {\bf 2}, 329 (2011).
\bibitem{CH} M. Campisi, P. H\"anggi, Entropy {\bf 13}, 2024 (2011).
%\bibitem{overlap}
%{\color{red}A quantitative measure for the overlap is given by $\Phi=\beta^{-1} \int d w \rho_f(w) \rho_b(-w)$. For the Gaussian work distribution (\ref{gauss}) one finds $\Phi = \exp [-(\beta \sigma)^2/2]/(2 \pi^{1/2} \beta \sigma)$ which become exponentially small for pronouncedly irreversible processes.}
\bibitem{jarzynski3}
C. Jarzynski, Phys. Rev. E {\bf 56}, 5018 (1997).
\bibitem{dhar}
A. Dhar, Phys. Rev. E {\bf 71}, 036126 (2005).


\end{thebibliography}
\end{document}